\documentclass[aps,twocolumn,pra,tightenlines,floatfix,showpacs]{revtex4-1}
\usepackage{graphicx}
\usepackage{amsmath}
\usepackage{amssymb}
\usepackage{times}
\usepackage[english]{babel}

\begin{document}

\title{Spatially varying interactions induced in ultra-cold atoms by optical Feshbach resonance}
\author{Chih-Chun Chien}

\affiliation{Theoretical Division, Los Alamos National Laboratory, MS B213, Los Alamos, NM 87545, USA}

\date{\today}

\begin{abstract}
Optical Feshbach resonance is capable of inducing spatially varying interactions in ultra-cold atoms. Its applications to pancake-shaped clouds of bosons and fermions enable one to study several fresh phenomena. We examine possibilities of inducing counter-intuitive structures such as creating a superfluid enclave inside a Mott insulator for bosons and a normal-gas core enclosed by a superfluid shell for fermions. We discuss feasible experimental setups and signatures of those interesting structures, which  can be very different from common structures observed in experiments so far. While a superfluid enclave in a Mott insulator can be useful for constructing atomic devices for atomtronics, the properties of the superconducting islands observed in scanning-tunneling microscopy of 
heavily underdoped high-temperature superconductors may be studied with cold Fermi gases with spatially varying attraction.
\end{abstract}


\maketitle

\section{Introduction}
Recent progress of experiments using ultra-cold atoms allows for simulations of complex quantum systems which can be formidable challenges to conventional setups \cite{PethickBEC,GeorgesReview}. Tunable atomic interactions and trapping potentials have brought us exciting phenomena such as BCS-Bose-Einstein condensation (BEC) crossover of fermionic superfluids and Mott insulator-superfluid transition of bosons \cite{PethickBEC}. More exotic phenomena such as heteronuclear superfluids \cite{Yip2mass} or $d$-wave superfluids \cite{Demler2} closely related to  nuclear physics or high-temperature superconductors (HTSCs) have been proposed. Here we show that recent development of optical Feshbach resonance (OFR) \cite{old_OFR_1,old_OFR_2,OFR1,bosonmodulation} which can control the scattering length in a selected region will bring another advantage of using ultra-cold atoms to demonstrate even more exciting quantum phenomena induced by controllable spatially varying interactions. Moreover, we argue that this technique is a key component in providing crucial links to other fields such as quantum simulations of electronic devices and HTSCs.

Atomtronics \cite{Atomtronics1}, whose goal is to simulate electronic devices using ultra-cold atoms, is a joint venture of several fields. A major challenge in atomtronics is to produce a designed pattern of atomic devices. In addition to engineered potentials, one can take advantage of spatially varying interactions, which can induce a superfluid region inside an insulator or vice versa as we will demonstrate here. A combination of the two techniques, spatially varying potential and interactions, will help realize  integrated atomic circuits, the analog of integrate circuits of electronics. 

An important property of spatially varying interactions generated by OFR is its capability of locally tuning the local density for a system in equilibrium. For bosons in the ground state this can be seen from Gross-Pitaevskii equation or Bose-Hubbard model, where the local chemical potential (defined as the difference between the equilibrium chemical potential and the local confining potential) depends on the local interaction strength \cite{PethickBEC}. For fermions with attractive interactions one can understand this from BCS-Leggett theory of BCS-BEC crossover of Fermi gases \cite{Leggett}, where the fermion chemical potential $\mu_f$ decreases monotonically as the attractive interaction increases.

For a trapped atomic gas, the local density approximation (LDA) suggests that the chemical potential has to balance the trap potential. In the presence of spatially varying interactions, the local chemical potential includes effects from both trap potential and the controlled interactions. As a consequence, the density profiles can change dramatically when the strength of interactions changes from one region to another.  

A major goal of this paper is to show that counter-intuitive structures can arise in well studied systems such as Bose-Hubbard model and BCS-BEC crossover. For example, a normal Fermi gas can coexist with a Fermi superfluid, but they have to arrange themselves to balance $\mu_f$ so that there is no net mass current. For a Fermi gas with uniform attractive interactions in a harmonic trap, the superfluid phase resides in the center of the trap, where the density is higher, to take advantage of its condensation energy. In contrast, by suitably shielding the attractive interactions at the trap center, it is possible to have a ground state with a normal-gas core encircled by a superfluid shell. We note that for equal-mass fermions with uniform interactions, the superfluid always resides at the trap center even in the presence of population imbalance \cite{Yip2mass,ChienPRL}. An inverted structure has been proposed in the ground state of heteronuclear Fermi gases with strongly attractive interactions \cite{Yip2mass} but has not been observed in experiments yet. Here we argue that a normal-gas core can exist in the ground state without introducing two different masses of fermions with mass imbalance if one can spatially vary the interactions using OFR. 

Another important motivation for implementing tunable spatially varying interactions comes from the fact that many interesting condensed matter systems are intrinsically inhomogeneous and the effective interactions thus change spatially. For example, superconducting islands immersed in a non-superconducting background have been reported in scanning-tunneling microscopy images of heavily underdoped cuprate HTSCs \cite{GomesSTM}. To help understand the mechanism of HTSCs, experiments on ultra-cold atoms should include such inhomogeneity effects and study how significantly those effects suppress the transition temperature.

There are other techniques for manipulating localized regions in an atomic cloud. For example, a spatial light modulator (SLM) \cite{SLM2003} can generate time-dependent patterns of trapping potentials for atoms so it is usually used as optical tweezers for atoms. There are experiments demonstrating splitting of BEC \cite{SLMBEC} or moving an atomic cloud \cite{SLMguide}. Thus SLMs are more suitable for studying dynamic properties. In this paper we focus on static configurations that can be generated by OFR, but together with the SLM one may further study dynamic phenomena.

Before presenting our theoretical studies, we briefly summarize relevant experimental studies on OFR. In bosonic BEC of $^{174}$Yb it has been demonstrated experimentally \cite{bosonmodulation} that a modulation of the scattering length with alternating regions of width $\sim 278$nm  can be generated by OFR. Such technique can be applied to fermionic $^{171}$Yb and $^{173}$Yb as well, which have been cooled down to $0.46T_F$ and $0.54T_F$ \cite{SU12}. Those temperatures are not far from the superfluid transition temperature of unitary Fermi gases, which is estimated to be $0.157T_F$ \cite{SalomonFL}. Moreover, one can introduce bosonic $^{172}$Yb or $^{174}$Yb into the fermionic system and the boson-fermion or boson-boson interactions can all be tuned by OFR \cite{OFRbosonfermion}. Adding to the excitement is the successful realization of OFR in bosonic $^{88}$Sr reported in Ref.~\cite{ThermoOFR}. Thus our predictions not only provide future directions but also are readily testable as experiments progress.

\section{Superfluid enclaves in a Mott insulator}
\begin{figure}
  \includegraphics[width=3.4in,clip] {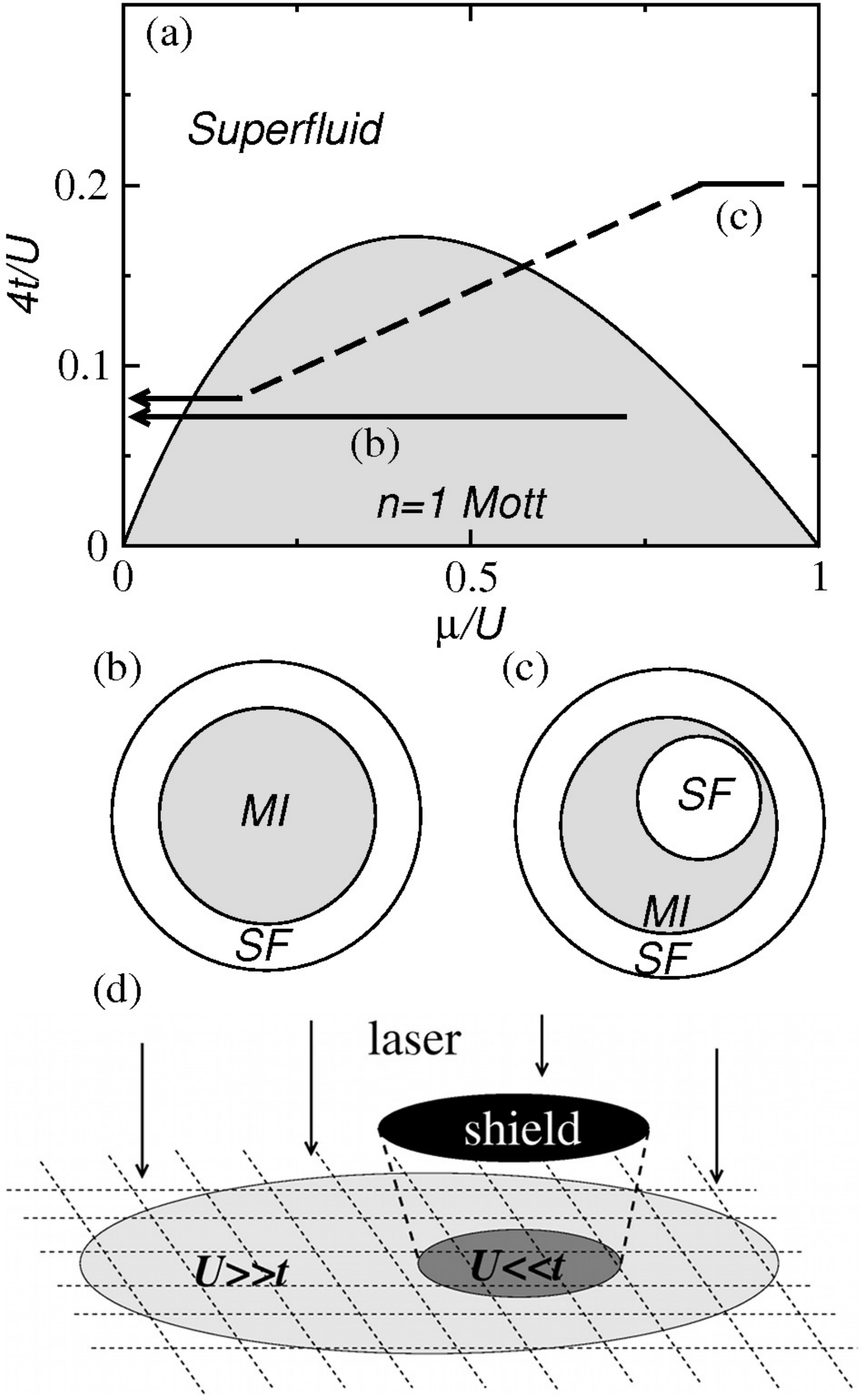}
  \caption{The phase diagram of 2D MI-superfluid transition (a). The two paths (b) and (c) show two possible phase structures from the trap center to the edge in a harmonic trap following LDA with $\mu(r)=\mu(r=0)-V_{trap}(r)$. For uniform interactions, a MI core is enclosed by a superfluid (SF) as shown in (b), which can be understood by following the path "(b)". When a shield blocks the enhancement of repulsion induced by OFR, the shielded region remains a SF, as shown in (c). Due to a sudden change in $U$ and continuity in $\mu$, there are jumps in both directions in (c) \cite{diffraction_note}. A proposed experimental setup is shown in (d), where we emphasize that the shield needs not be above the trap center. The structure of (c) can be understood by following the path "(c)". Here $t$ and $U$ are the hopping coefficient and onsite repulsion. Possible diffraction patterns inside the shadowed region are not shown.}
\label{fig:MISF}
\end{figure}

We first show how spatially varying interactions can induce a superfluid enclave inside an insulating region. Here we consider Bose-Hubbard model (see Ref.~\cite{PethickBEC} for a review) describing bosons on optical lattices in quasi two dimension (2D). The Hamiltonian is
\begin{equation}
H_{BH}=-t\sum_{\langle ij\rangle}b^{\dagger}_{i}b_{j}-\mu\sum_{i}n_{i}+\frac{U}{2}\sum_{i}n_{i}(n_{i}-1).
\end{equation}
Here $b_{i}$ ($b^{\dagger}_{i}$) is the boson annihilation (creation) operator, $\langle ij\rangle$ denotes nearest neighbors, $t$ and $U$ are the hopping coefficient and onsite repulsive coupling constant. There is also a background harmonic trap potential $V_{trap}=m\omega_{tr}^{2}r^2/2$, where $\omega_{tr}$ is the trap frequency.
The Mott insulator (MI) - superfluid transition in quasi 2D has been demonstrated experimentally \cite{CL2DMott}. When the onsite repulsion $U$ is strong (weak), bosons localize (delocalize) and form a Mott insulator (superfluid). Figure~\ref{fig:MISF}(a) shows the phase diagram of the $n=1$ MI and the superfluid phase close to it (see Ref.~\cite{PethickBEC} for details), where $n$ is the filling factor. 

In the presence of a harmonic trap, the chemical potential decreases from the trap center to the edge as shown by the path labeled (b) on the phase diagram. Thus the $n=1$ MI resides at the trap center enclosed by a superfluid at the edge. The structure is schematically shown in Fig.~\ref{fig:MISF}(b). Now we consider tuning $U$ using OFR  because $U$  is controlled by the scattering length $a$ \cite{Demler2}, and one expects to see a similar structure. However, one has the advantage of locally tuning the interactions using OFR. 

We consider a quasi 2D optical lattice loaded with $^{174}$Yb though other bosonic isotopes will also work. One can increase $U$ by using OFR and induce a similar structure as shown in Fig.~\ref{fig:MISF}(b). If one shields a small region of the MI so that instead of strong repulsion, bosons in the shielded region interact with each other with a background scattering length $a_{bg}=5.55$nm \cite{bosonmodulation}, which should correspond to a small $U$ for a lattice with moderate depth.  Thus the shielded region will be a superfluid. The setup is shown in Fig.~\ref{fig:MISF}(d). The path labeled (c) on the phase diagram shows how the chemical potential and interaction strength vary as one goes from the shielded region toward the trap edge. The structure of the atomic cloud under such an engineered interaction pattern is schematically shown in Fig.~\ref{fig:MISF}(c).

It is known that one can generate a superfluid region inside a MI core simply by increasing the filling factor. In that way the filling factor exceeds $n=1$ at the trap center so a superfluid core emerges. However, the advantage of using spatially varying interactions to induce a superfluid enclave in the MI core is that the enclave can, in principle, be induced \textit{anywhere} in the MI core, not just at the trap center. We deliberately demonstrate this on Fig.~\ref{fig:MISF}(c). Moreover, one can induce more than one superfluid enclaves in the MI core as long as the resolution allows. With suitably designed masks, one can expect that this technique will help realize "printed circuits" of atomic devices. We emphasize that similar phenomena can be studied using fermions as well. We remark that diffractive effects may induce further inhomogeneity effects inside the shadowed region such as Arago spot or diffractive oscillations. Experimentally one may use a mask with a corrugated edge or a combination of masks and lens to minimize diffraction.

So far we consider how to induce a superfluid enclave inside a MI core. One can as well induce a MI enclave inside a superfluid. One possible setup is a weakly-interacting superfluid with a scattering length $a_{bg}$ and filling factor slightly above $n=1$. Then by imposing OFR in localized regions the repulsion $U$ can be increased and when it exceeds the critical value, MI enclaves are expected to emerge. Since the density profile of a quasi 2D Bose gas in optical lattices can be measured quite accurately \cite{CL2DMott}, one should be able to identify those enclaves. In real experiments, however,  there should be a transient region (or a finite-width wall) between the MI and the induced superfluid region. The width is estimated to be of order of $\xi_c=1/\sqrt{8\pi n a_{bg}}$, the coherence length of the condensate inside the enclave, because it determines the healing length \cite{PethickBEC}. From the parameters shown in Ref.~\cite{bosonmodulation}, we estimate that $\xi_c\approx 360$nm in the central region, which is reasonably small compare to the Thomas-Fermi radius of that setup, which is $5.5\mu$m.

One future application may be a superfluid-MI-superfluid junction constructed by bringing the superfluid enclave close to the outer superfluid ring for  tunneling phenomena. Such a tunneling junction will complement the Josephson junction in the absence of optical lattices studied in Ref.~\cite{boson_Josephson_Junction,JJunction}.

\section{Inverted structures of Fermi gases}
One major difference between bosonic and fermion systems is that in the absence of lattice potentials, the low-temperature phase of a Bose gas is always a superfluid. One has to go above the BEC temperature to observe a normal phase. To have coexistence of different phases at low $T$ thus requires lattice potentials to induce an insulating (MI) phase. In contrast, for a Fermi gas with attractive interactions, the superfluid transition temperature decreases exponentially with the coupling constant in the weakly attractive regime \cite{PethickBEC}. For reasonably low temperature there is a corresponding critical interaction strength below which the system is a normal Fermi gas. Therefore one can induce interesting structures in harmonically trapped Fermi gases without optical lattices.

\begin{figure}
  \includegraphics[width=3.4in,clip] {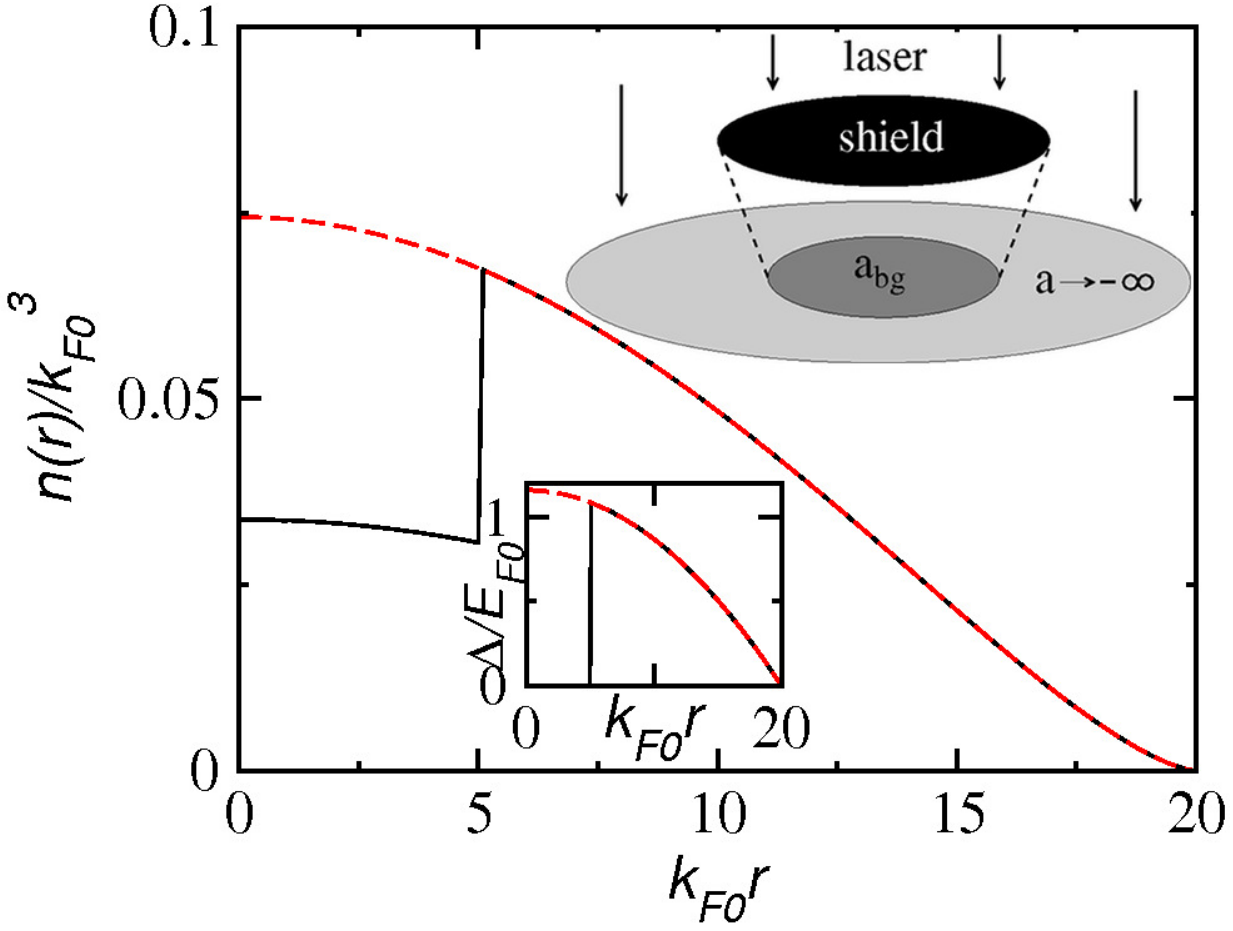}
  \caption{(Color online) The trap profiles of $n$ and $\Delta$ (lower inset). Solid/dashed lines correspond to spatially varying/uniform interactions. Here $E_{F0}$ and $k_{F0}$ are the Fermi energy and momentum at the trap center. The proposed experimental setup is shown in the upper inset. The atomic cloud is a unitary Fermi gas induced by OFR, but the central region is shielded and remains a normal gas. Possible diffraction patterns inside the shadowed region are not shown.}
\label{fig:2}
\end{figure}  
One counter-intuitive case of using OFR in Fermi gases is to induce a normal-gas region inside a fermionic superfluid. 
In the inset of Figure~\ref{fig:2} we show a setup that can generate such a structure. It is similar to the setup of Fig.~\ref{fig:MISF} (d) except there is no optical lattice. We consider $^{171}$Yb with background scattering length $a_{bg}=-0.15$nm \cite{SU12} and assume the cloud is at unitarity induced by OFR, i.e., $a\rightarrow -\infty$. Nonetheless, a shield blocks OFR at the trap center so the atoms there interact with each other via $a_{bg}$, which corresponds to extremely small attraction \footnote{$k_{F}a_{bg}\approx -1.5\times 10^{-3}$ at the trap center from Ref.\cite{SU12} so $T_c$ is well below experimentally accessible temperatures so far.}. At low but finite $T$ (say $T=0.05T_F$, where $T_F$ is the local Fermi temperature), the shielded region is a normal gas while most of the rest of the cloud is a unitary Fermi gas, which is a superfluid.

For demonstration, we consider the situation where $a=0^{-}$ for $r<r_{c}$ and $a=-\infty$ for $r>r_{c}$. We choose $r_{c}=(1/4)R_{TF}$ and $\hbar\omega_{r}=0.1 E_{F0}$, where $\omega_{r}$ is the radial trap frequency and $R_{TF}$ is the Thomas-Fermi radius. Since the unitary limit is around the place where the transition temperature shows a maximum in BCS-BEC crossover \cite{PethickBEC}, this choice makes the shell structure easier to be realized in experiments. We begin with the ground state close to $T=0$.

In the presence of a harmonic trap, one can use LDA and the BCS-Leggett theory to obtain the following equations of states \cite{Leggett,PethickBEC} 
\begin{eqnarray}
n(r)&=&\sum_{\bf k}\left(1-\frac{\epsilon_{k}-\mu_f(r)}{E_{k}}\right), \nonumber \\
\frac{m}{4\pi\hbar^2 a}&=&\sum_{\bf k}\left(\frac{1}{2\epsilon_{k}}-\frac{1}{2E_{k}}\right).
\end{eqnarray}
Here $E_{k}=\sqrt{(\epsilon_{k}-\mu_{f}(r))^2+\Delta(r)^2}$ and $\epsilon_{k}=\hbar^2 k^2/(2m)$. The trap profiles of density and order parameter $\Delta$ of the Fermi gas are shown in Figure~\ref{fig:2}. One immediately sees that in order to establish equilibrium with a continuous $\mu_f$, the density in the superfluid phase has to jump in order to balance the diffusion on both sides. Such discontinuities in density profiles have been shown in the phase-separated structure of polarized Fermi gases due to the same mechanism \cite{ChienPRL}, i.e., the balance of $\mu_f$, but a normal-gas core cannot be realized if the mass of the two species are the same.

The jump in the density profile serves as direct evidence that a superfluid shell encircles a normal-gas core for an unpolarized Fermi gas with equal mass. One can also confirm the coexistence of a superfluid and a normal gas by measuring the total energy $E$ of the system from the cloud size \cite{ThermoScience}. The condensation energy from the superfluid at the trap edge will make $E$ smaller than that of a cloud consisting of normal gas with the same total particle number.  To contrast our predictions with the case with a uniform interaction, we also show in Figure~\ref{fig:2} the profiles of $n$ and $\Delta$ of a trapped unitary Fermi gas obtained from the same profile of chemical potential. There is no jump in the profiles and the whole cloud is a superfluid.

We remark that the sharp discontinuity in the trap profiles of $n$ and $\Delta$ is also an artifact of LDA and will be broadened due to the kinetic energy terms not considered in LDA. However, finite jumps in the density should still be observable based on the experimental observations of phase-separated structures in polarized Fermi gases \cite{Zwierlein2006}. Again we can estimate the width of the transient region by the coherence length of the superfluid phase, whose magnitude is estimated from the BCS formula $\xi_{BCS}=\hbar v_{F}/\Delta$ \cite{PethickBEC}, where $v_F$ is the Fermi velocity. From the parameters of Ref.~\cite{SU12}, $\xi_{BCS}\approx 210$nm, which is reasonably small compared to the radius of the cloud ($\approx 30\mu$m). We emphasize that it is the jump in the density that gives an important signature of coexistence of different phases and the sharpness of the boundary is less relevant. The validity of using LDA and Leggett-BCS theory to describe phase-separated structures of polarized Fermi gases \cite{ChienPRL} also support the validity of using the same method for unpolarized Fermi gases with spatially-varying interactions.

At finite $T$, the edge of the cloud with low density will become a normal gas due to thermal excitations. This applies to both the case with spatially varying interactions as well as the case with a uniform interaction strength. As a consequence, a cloud with a uniform interaction has a superfluid core encircled by a normal gas shell at the trap edge. In contrast, the structure of a cloud with spatially varying interactions as shown in Fig.~\ref{fig:2} will be a three-layer one with a normal-gas core due to the absence of OFR, a superfluid shell, and another normal gas shell due to thermal excitations. By introducing more complicated patterns of spatially varying interactions, one may induce structures with more alternating layers, which can be identified from jumps in the density profile.

There is another interesting property of such an inverted shell structure. When the cloud is under rotation around the center, the superfluid can simply circulate around the normal-gas core, which is like a giant vortex core. Since angular momentum can be carried in this way, there should be no vortex in the superfluid region. The absence of any vortex when the cloud is under rotation thus can serve as another evidence that the core consists of a normal gas.

An important contribution of realizing ultra-cold Fermi gases with spatially varying interactions is to provide information of how superfluid islands can be stabilized in a quasi-2D norma-gas matrix by inhomogeneous interaction strength. For fermionic clouds such inhomogeneity can produce modulations in the density as well as modulations in the order parameter. Moreover, the superfluid regions should have higher density similar to that shown in Fig.~\ref{fig:2} so they are really "islands" standing on the normal background. This  setup could be useful in estimating how imbalanced the interactions may be for the observed supeconducting islands in underdoped HTSCs as shown in Ref.~\cite{GomesSTM}.

When fermions are loaded into optical lattices, more interesting phases are expected to emerge, including fermionic Mott insulator, magnetic phases, $d$-wave superfluid, etc \cite{Demler2}. Since those phases depend on the strength of interactions, with carefully designed masks for OFR laser, one should observe many more interesting structures which may not be possible in conventional condensed matter systems.

\section{Summary}
In both bosonic and fermionic systems we present feasible experimental setups for generating counter-intuitive structures, signatures that can identify the targeted phenomena, and experimental techniques that can measure those signatures. We propose that OFR may enable ultra-cold atomic gases to simulate electronics, heavily underdoped cuprate superconductors, and more. This technique complements others for ultra-cold atoms such as optical lattices, SLM, polarized Fermi gases, etc. One can envision that combinations of those techniques could demonstrate more exciting macroscopic quantum phenomena.

In this paper we focus on counter-intuitive phenomena in equilibrium because they may be realized with minimal adjustment to current experimental setups such as those described in Refs.~\cite{bosonmodulation,SU12}. We point out that spatially varying interaction induced by OFR also has great potential in the study of quantum transport. For example, one has seen that local chemical potential can be tuned by OFR. A more general definition of current is the \textit{electrochemical} current, which is driven by electric field (for charged systems) and/or chemical potential. With time-dependent interactions, one can produce chemical potential imbalance and a mass current will flow. One advantage of using Yb isotopes and OFR for studying quantum transport is the tunable interactions between different isotopes \cite{OFRbosonfermion}. A possible direction may be to co-trap different species  (bosons and fermions) and simulate the suppression of the current by impurities.

\section*{Acknowledgement}
The author thanks M. Zwolak, Y. Takahashi, and B. Damski for useful discussions. This work is supported by U.S. DOE through the LANL/LDRD Program.

\bibliographystyle{apsrev4-1}

\begin{thebibliography}{26}%
\makeatletter
\providecommand \@ifxundefined [1]{%
 \@ifx{#1\undefined}
}%
\providecommand \@ifnum [1]{%
 \ifnum #1\expandafter \@firstoftwo
 \else \expandafter \@secondoftwo
 \fi
}%
\providecommand \@ifx [1]{%
 \ifx #1\expandafter \@firstoftwo
 \else \expandafter \@secondoftwo
 \fi
}%
\providecommand \natexlab [1]{#1}%
\providecommand \enquote  [1]{``#1''}%
\providecommand \bibnamefont  [1]{#1}%
\providecommand \bibfnamefont [1]{#1}%
\providecommand \citenamefont [1]{#1}%
\providecommand \href@noop [0]{\@secondoftwo}%
\providecommand \href [0]{\begingroup \@sanitize@url \@href}%
\providecommand \@href[1]{\@@startlink{#1}\@@href}%
\providecommand \@@href[1]{\endgroup#1\@@endlink}%
\providecommand \@sanitize@url [0]{\catcode `\\12\catcode `\$12\catcode
  `\&12\catcode `\#12\catcode `\^12\catcode `\_12\catcode `\%12\relax}%
\providecommand \@@startlink[1]{}%
\providecommand \@@endlink[0]{}%
\providecommand \url  [0]{\begingroup\@sanitize@url \@url }%
\providecommand \@url [1]{\endgroup\@href {#1}{\urlprefix }}%
\providecommand \urlprefix  [0]{URL }%
\providecommand \Eprint [0]{\href }%
\@ifxundefined \urlstyle {%
  \providecommand \doi  [0]{\begingroup \@sanitize@url \@doi}%
  \providecommand \@doi [1]{\endgroup \@@startlink {\doibase
  #1}doi:\discretionary {}{}{}#1\@@endlink }%
}{%
  \providecommand \doi  [0]{doi:\discretionary{}{}{}\begingroup
  \urlstyle{rm}\Url }%
}%
\providecommand \doibase [0]{http://dx.doi.org/}%
\providecommand \Doi [0]{\begingroup \@sanitize@url \@Doi }%
\providecommand \@Doi  [1]{\endgroup\@@startlink{\doibase#1}\@@Doi}%
\providecommand \@@Doi [1]{#1\@@endlink}%
\providecommand \selectlanguage [0]{\@gobble}%
\providecommand \bibinfo  [0]{\@secondoftwo}%
\providecommand \bibfield  [0]{\@secondoftwo}%
\providecommand \translation [1]{[#1]}%
\providecommand \BibitemOpen [0]{}%
\providecommand \bibitemStop [0]{}%
\providecommand \bibitemNoStop [0]{.\EOS\space}%
\providecommand \EOS [0]{\spacefactor3000\relax}%
\providecommand \BibitemShut  [1]{\csname bibitem#1\endcsname}%
\bibitem [{\citenamefont {Pethick}\ and\ \citenamefont
  {Smith}(2008)}]{PethickBEC}%
  \BibitemOpen
  \bibfield  {author} {\bibinfo {author} {\bibfnamefont {C.~J.}\ \bibnamefont
  {Pethick}}\ and\ \bibinfo {author} {\bibfnamefont {H.}~\bibnamefont
  {Smith}},\ }\href@noop {} {\emph {\bibinfo {title} {Bose-Einstein
  Condensation in Dilute Gases}}},\ \bibinfo {edition} {2nd}\ ed.\ (\bibinfo
  {publisher} {Cambridge University Press},\ \bibinfo {address} {Cambridge,
  UK},\ \bibinfo {year} {2008})\BibitemShut {NoStop}%
\bibitem [{\citenamefont {Georges}(2007)}]{GeorgesReview}%
  \BibitemOpen
  \bibfield  {author} {\bibinfo {author} {\bibfnamefont {A.}~\bibnamefont
  {Georges}},\ }in\ \href@noop {} {\emph {\bibinfo {booktitle} {Ultra-cold
  Fermi Gases}}},\ \bibinfo {editor} {edited by\ \bibinfo {editor}
  {\bibfnamefont {M.}~\bibnamefont {Inguscio}}, \bibinfo {editor}
  {\bibfnamefont {W.}~\bibnamefont {Ketterle}}, \ and\ \bibinfo {editor}
  {\bibfnamefont {C.}~\bibnamefont {Salomon}}}\ (\bibinfo  {publisher} {Italian
  physical society},\ \bibinfo {year} {2007})\ p.\ \bibinfo {pages} {477},\
  \bibinfo {note} {arXiv:cond-mat/0702122}\BibitemShut {NoStop}%
\bibitem [{Yip()}]{Yip2mass}%
  \BibitemOpen
  \href@noop {} {}\bibinfo {note} {S. T. Wu, C. H. Pao, and S. K. Yip, \prb
  \textbf{74}, 224504 (2006); C. H. Pao, S. T. Wu, and S. K. Yip, \pra
  \textbf{76}, 053621 (2007).}\BibitemShut {Stop}%
\bibitem [{\citenamefont {Hofstetter}\ \emph {et~al.}(2002)\citenamefont
  {Hofstetter}, \citenamefont {Cirac}, \citenamefont {Zoller}, \citenamefont
  {Demler},\ and\ \citenamefont {Lukin}}]{Demler2}%
  \BibitemOpen
  \bibfield  {author} {\bibinfo {author} {\bibfnamefont {W.}~\bibnamefont
  {Hofstetter}}, \bibinfo {author} {\bibfnamefont {J.~I.}\ \bibnamefont
  {Cirac}}, \bibinfo {author} {\bibfnamefont {P.}~\bibnamefont {Zoller}},
  \bibinfo {author} {\bibfnamefont {E.}~\bibnamefont {Demler}}, \ and\ \bibinfo
  {author} {\bibfnamefont {M.~D.}\ \bibnamefont {Lukin}},\ }\href@noop {}
  {\bibfield  {journal} {\bibinfo  {journal} {Phys. Rev. Lett.},\ }\textbf
  {\bibinfo {volume} {89}},\ \bibinfo {pages} {220407} (\bibinfo {year}
  {2002})}\BibitemShut {NoStop}%
\bibitem [{\citenamefont {Fatemi}\ \emph {et~al.}(2000)\citenamefont {Fatemi},
  \citenamefont {Jones},\ and\ \citenamefont {Lett}}]{old_OFR_1}%
  \BibitemOpen
  \bibfield  {author} {\bibinfo {author} {\bibfnamefont {F.~K.}\ \bibnamefont
  {Fatemi}}, \bibinfo {author} {\bibfnamefont {K.~M.}\ \bibnamefont {Jones}}, \
  and\ \bibinfo {author} {\bibfnamefont {P.~D.}\ \bibnamefont {Lett}},\
  }\href@noop {} {\bibfield  {journal} {\bibinfo  {journal} {Phys. Rev.
  Lett.},\ }\textbf {\bibinfo {volume} {85}},\ \bibinfo {pages} {4462}
  (\bibinfo {year} {2000})}\BibitemShut {NoStop}%
\bibitem [{\citenamefont {Theis}\ \emph {et~al.}(2004)\citenamefont {Theis},
  \citenamefont {Thalhammer}, \citenamefont {Winkler}, \citenamefont {Hellwig},
  \citenamefont {Ruff}, \citenamefont {Grimm},\ and\ \citenamefont
  {Hecker~Denschlag}}]{old_OFR_2}%
  \BibitemOpen
  \bibfield  {author} {\bibinfo {author} {\bibfnamefont {M.}~\bibnamefont
  {Theis}}, \bibinfo {author} {\bibfnamefont {G.}~\bibnamefont {Thalhammer}},
  \bibinfo {author} {\bibfnamefont {K.}~\bibnamefont {Winkler}}, \bibinfo
  {author} {\bibfnamefont {M.}~\bibnamefont {Hellwig}}, \bibinfo {author}
  {\bibfnamefont {G.}~\bibnamefont {Ruff}}, \bibinfo {author} {\bibfnamefont
  {R.}~\bibnamefont {Grimm}}, \ and\ \bibinfo {author} {\bibfnamefont
  {J.}~\bibnamefont {Hecker~Denschlag}},\ }\href@noop {} {\bibfield  {journal}
  {\bibinfo  {journal} {Phys. Rev. Lett.},\ }\textbf {\bibinfo {volume} {93}},\
  \bibinfo {pages} {123001} (\bibinfo {year} {2004})}\BibitemShut {NoStop}%
\bibitem [{\citenamefont {Fedichev}\ \emph {et~al.}(1996)\citenamefont
  {Fedichev}, \citenamefont {Kagan}, \citenamefont {Shlyapnikov},\ and\
  \citenamefont {Walraven}}]{OFR1}%
  \BibitemOpen
  \bibfield  {author} {\bibinfo {author} {\bibfnamefont {P.~O.}\ \bibnamefont
  {Fedichev}}, \bibinfo {author} {\bibfnamefont {Y.}~\bibnamefont {Kagan}},
  \bibinfo {author} {\bibfnamefont {G.~V.}\ \bibnamefont {Shlyapnikov}}, \ and\
  \bibinfo {author} {\bibfnamefont {J.~T.~M.}\ \bibnamefont {Walraven}},\
  }\href@noop {} {\bibfield  {journal} {\bibinfo  {journal} {Phys. Rev.
  Lett.},\ }\textbf {\bibinfo {volume} {77}},\ \bibinfo {pages} {2913}
  (\bibinfo {year} {1996})}\BibitemShut {NoStop}%
\bibitem [{\citenamefont {Yamazaki}\ \emph {et~al.}(2010)\citenamefont
  {Yamazaki}, \citenamefont {Taie}, \citenamefont {Sugawa},\ and\ \citenamefont
  {Takahashi}}]{bosonmodulation}%
  \BibitemOpen
  \bibfield  {author} {\bibinfo {author} {\bibfnamefont {R.}~\bibnamefont
  {Yamazaki}}, \bibinfo {author} {\bibfnamefont {S.}~\bibnamefont {Taie}},
  \bibinfo {author} {\bibfnamefont {S.}~\bibnamefont {Sugawa}}, \ and\ \bibinfo
  {author} {\bibfnamefont {Y.}~\bibnamefont {Takahashi}},\ }\href@noop {}
  {\bibfield  {journal} {\bibinfo  {journal} {Phys. Rev. Lett.},\ }\textbf
  {\bibinfo {volume} {105}},\ \bibinfo {pages} {050405} (\bibinfo {year}
  {2010})}\BibitemShut {NoStop}%
\bibitem [{\citenamefont {Pepino}\ \emph {et~al.}(2009)\citenamefont {Pepino},
  \citenamefont {Cooper}, \citenamefont {Anderson},\ and\ \citenamefont
  {Holland}}]{Atomtronics1}%
  \BibitemOpen
  \bibfield  {author} {\bibinfo {author} {\bibfnamefont {R.~A.}\ \bibnamefont
  {Pepino}}, \bibinfo {author} {\bibfnamefont {J.}~\bibnamefont {Cooper}},
  \bibinfo {author} {\bibfnamefont {D.~Z.}\ \bibnamefont {Anderson}}, \ and\
  \bibinfo {author} {\bibfnamefont {M.~J.}\ \bibnamefont {Holland}},\
  }\href@noop {} {\bibfield  {journal} {\bibinfo  {journal} {Phys. Rev.
  Lett.},\ }\textbf {\bibinfo {volume} {103}},\ \bibinfo {pages} {140405}
  (\bibinfo {year} {2009})}\BibitemShut {NoStop}%
\bibitem [{\citenamefont {Leggett}(1980)}]{Leggett}%
  \BibitemOpen
  \bibfield  {author} {\bibinfo {author} {\bibfnamefont {A.~J.}\ \bibnamefont
  {Leggett}},\ }in\ \href@noop {} {\emph {\bibinfo {booktitle} {Modern Trends
  in the Theory of Condensed Matter}}}\ (\bibinfo  {publisher}
  {Springer-Verlag},\ \bibinfo {address} {Berlin},\ \bibinfo {year} {1980})\
  pp.\ \bibinfo {pages} {13--27}\BibitemShut {NoStop}%
\bibitem [{\citenamefont {Chien}\ \emph {et~al.}(2007)\citenamefont {Chien},
  \citenamefont {Chen}, \citenamefont {He},\ and\ \citenamefont
  {Levin}}]{ChienPRL}%
  \BibitemOpen
  \bibfield  {author} {\bibinfo {author} {\bibfnamefont {C.-C.}\ \bibnamefont
  {Chien}}, \bibinfo {author} {\bibfnamefont {Q.~J.}\ \bibnamefont {Chen}},
  \bibinfo {author} {\bibfnamefont {Y.}~\bibnamefont {He}}, \ and\ \bibinfo
  {author} {\bibfnamefont {K.}~\bibnamefont {Levin}},\ }\href@noop {}
  {\bibfield  {journal} {\bibinfo  {journal} {Phys. Rev. Lett.},\ }\textbf
  {\bibinfo {volume} {98}},\ \bibinfo {pages} {110404} (\bibinfo {year}
  {2007})}\BibitemShut {NoStop}%
\bibitem [{\citenamefont {Gomes}\ \emph {et~al.}(2007)\citenamefont {Gomes},
  \citenamefont {Pasupathy}, \citenamefont {Pushp}, \citenamefont {Ono},
  \citenamefont {Ando},\ and\ \citenamefont {Yazdani}}]{GomesSTM}%
  \BibitemOpen
  \bibfield  {author} {\bibinfo {author} {\bibfnamefont {K.~K.}\ \bibnamefont
  {Gomes}}, \bibinfo {author} {\bibfnamefont {A.~N.}\ \bibnamefont
  {Pasupathy}}, \bibinfo {author} {\bibfnamefont {A.}~\bibnamefont {Pushp}},
  \bibinfo {author} {\bibfnamefont {S.}~\bibnamefont {Ono}}, \bibinfo {author}
  {\bibfnamefont {Y.}~\bibnamefont {Ando}}, \ and\ \bibinfo {author}
  {\bibfnamefont {A.}~\bibnamefont {Yazdani}},\ }\href@noop {} {\bibfield
  {journal} {\bibinfo  {journal} {Nature},\ }\textbf {\bibinfo {volume}
  {447}},\ \bibinfo {pages} {569} (\bibinfo {year} {2007})}\BibitemShut
  {NoStop}%
\bibitem [{\citenamefont {McGloin}\ \emph {et~al.}(2003)\citenamefont
  {McGloin}, \citenamefont {Spalding}, \citenamefont {Melville}, \citenamefont
  {Sibbett},\ and\ \citenamefont {Dholakia}}]{SLM2003}%
  \BibitemOpen
  \bibfield  {author} {\bibinfo {author} {\bibfnamefont {D.}~\bibnamefont
  {McGloin}}, \bibinfo {author} {\bibfnamefont {G.~C.}\ \bibnamefont
  {Spalding}}, \bibinfo {author} {\bibfnamefont {H.}~\bibnamefont {Melville}},
  \bibinfo {author} {\bibfnamefont {W.}~\bibnamefont {Sibbett}}, \ and\
  \bibinfo {author} {\bibfnamefont {K.}~\bibnamefont {Dholakia}},\ }\href@noop
  {} {\bibfield  {journal} {\bibinfo  {journal} {Opt. Express},\ }\textbf
  {\bibinfo {volume} {11}},\ \bibinfo {pages} {158} (\bibinfo {year}
  {2003})}\BibitemShut {NoStop}%
\bibitem [{\citenamefont {Boyer}\ \emph {et~al.}(2006)\citenamefont {Boyer},
  \citenamefont {Godun}, \citenamefont {Smirne}, \citenamefont {Cassettari},
  \citenamefont {Chandrashekar}, \citenamefont {Deb}, \citenamefont {Laczik},\
  and\ \citenamefont {Foot}}]{SLMBEC}%
  \BibitemOpen
  \bibfield  {author} {\bibinfo {author} {\bibfnamefont {V.}~\bibnamefont
  {Boyer}}, \bibinfo {author} {\bibfnamefont {R.~M.}\ \bibnamefont {Godun}},
  \bibinfo {author} {\bibfnamefont {G.}~\bibnamefont {Smirne}}, \bibinfo
  {author} {\bibfnamefont {D.}~\bibnamefont {Cassettari}}, \bibinfo {author}
  {\bibfnamefont {C.~M.}\ \bibnamefont {Chandrashekar}}, \bibinfo {author}
  {\bibfnamefont {A.~B.}\ \bibnamefont {Deb}}, \bibinfo {author} {\bibfnamefont
  {Z.~J.}\ \bibnamefont {Laczik}}, \ and\ \bibinfo {author} {\bibfnamefont
  {C.~J.}\ \bibnamefont {Foot}},\ }\href@noop {} {\bibfield  {journal}
  {\bibinfo  {journal} {Phys. Rev. A},\ }\textbf {\bibinfo {volume} {73}},\
  \bibinfo {pages} {031402(R)} (\bibinfo {year} {2006})}\BibitemShut {NoStop}%
\bibitem [{\citenamefont {Fatemi}\ \emph {et~al.}(2007)\citenamefont {Fatemi},
  \citenamefont {Bashkansky},\ and\ \citenamefont {Dutton}}]{SLMguide}%
  \BibitemOpen
  \bibfield  {author} {\bibinfo {author} {\bibfnamefont {F.~K.}\ \bibnamefont
  {Fatemi}}, \bibinfo {author} {\bibfnamefont {M.}~\bibnamefont {Bashkansky}},
  \ and\ \bibinfo {author} {\bibfnamefont {Z.}~\bibnamefont {Dutton}},\
  }\href@noop {} {\bibfield  {journal} {\bibinfo  {journal} {Opt. Express},\
  }\textbf {\bibinfo {volume} {15}},\ \bibinfo {pages} {3589} (\bibinfo {year}
  {2007})}\BibitemShut {NoStop}%
\bibitem [{\citenamefont {Taie}\ \emph {et~al.}(2010)\citenamefont {Taie},
  \citenamefont {Takasu}, \citenamefont {Sugawa}, \citenamefont {Yamazaki},
  \citenamefont {Tsujimoto}, \citenamefont {Murakami},\ and\ \citenamefont
  {Takahashi}}]{SU12}%
  \BibitemOpen
  \bibfield  {author} {\bibinfo {author} {\bibfnamefont {S.}~\bibnamefont
  {Taie}}, \bibinfo {author} {\bibfnamefont {Y.}~\bibnamefont {Takasu}},
  \bibinfo {author} {\bibfnamefont {S.}~\bibnamefont {Sugawa}}, \bibinfo
  {author} {\bibfnamefont {R.}~\bibnamefont {Yamazaki}}, \bibinfo {author}
  {\bibfnamefont {T.}~\bibnamefont {Tsujimoto}}, \bibinfo {author}
  {\bibfnamefont {R.}~\bibnamefont {Murakami}}, \ and\ \bibinfo {author}
  {\bibfnamefont {Y.}~\bibnamefont {Takahashi}},\ }\href@noop {} {\bibfield
  {journal} {\bibinfo  {journal} {Phys. Rev. Lett.},\ }\textbf {\bibinfo
  {volume} {105}},\ \bibinfo {pages} {190401} (\bibinfo {year}
  {2010})}\BibitemShut {NoStop}%
\bibitem [{\citenamefont {Nascimbene}\ \emph {et~al.}(2010)\citenamefont
  {Nascimbene}, \citenamefont {Navon}, \citenamefont {J}, \citenamefont
  {Chevy},\ and\ \citenamefont {Salomon}}]{SalomonFL}%
  \BibitemOpen
  \bibfield  {author} {\bibinfo {author} {\bibfnamefont {S.}~\bibnamefont
  {Nascimbene}}, \bibinfo {author} {\bibfnamefont {N.}~\bibnamefont {Navon}},
  \bibinfo {author} {\bibfnamefont {J.~K.}\ \bibnamefont {J}}, \bibinfo
  {author} {\bibfnamefont {F.}~\bibnamefont {Chevy}}, \ and\ \bibinfo {author}
  {\bibfnamefont {C.}~\bibnamefont {Salomon}},\ }\href@noop {} {\bibfield
  {journal} {\bibinfo  {journal} {Nature},\ }\textbf {\bibinfo {volume}
  {463}},\ \bibinfo {pages} {1057} (\bibinfo {year} {2010})}\BibitemShut
  {NoStop}%
\bibitem [{\citenamefont {Fukuhara}\ \emph {et~al.}(2009)\citenamefont
  {Fukuhara}, \citenamefont {Sugawa}, \citenamefont {Takasu},\ and\
  \citenamefont {Takahashi}}]{OFRbosonfermion}%
  \BibitemOpen
  \bibfield  {author} {\bibinfo {author} {\bibfnamefont {T.}~\bibnamefont
  {Fukuhara}}, \bibinfo {author} {\bibfnamefont {S.}~\bibnamefont {Sugawa}},
  \bibinfo {author} {\bibfnamefont {Y.}~\bibnamefont {Takasu}}, \ and\ \bibinfo
  {author} {\bibfnamefont {Y.}~\bibnamefont {Takahashi}},\ }\href@noop {}
  {\bibfield  {journal} {\bibinfo  {journal} {Phys. Rev. A},\ }\textbf
  {\bibinfo {volume} {79}},\ \bibinfo {pages} {021601} (\bibinfo {year}
  {2009})}\BibitemShut {NoStop}%
\bibitem [{\citenamefont {Blatt}\ \emph {et~al.}()\citenamefont {Blatt},
  \citenamefont {Nicholson}, \citenamefont {Bloom}, \citenamefont {Williams},
  \citenamefont {Thomsen}, \citenamefont {Julienne},\ and\ \citenamefont
  {Ye}}]{ThermoOFR}%
  \BibitemOpen
  \bibfield  {author} {\bibinfo {author} {\bibfnamefont {S.}~\bibnamefont
  {Blatt}}, \bibinfo {author} {\bibfnamefont {T.~L.}\ \bibnamefont
  {Nicholson}}, \bibinfo {author} {\bibfnamefont {B.~J.}\ \bibnamefont
  {Bloom}}, \bibinfo {author} {\bibfnamefont {J.~R.}\ \bibnamefont {Williams}},
  \bibinfo {author} {\bibfnamefont {J.~W.}\ \bibnamefont {Thomsen}}, \bibinfo
  {author} {\bibfnamefont {P.~S.}\ \bibnamefont {Julienne}}, \ and\ \bibinfo
  {author} {\bibfnamefont {J.}~\bibnamefont {Ye}},\ }\href@noop {} {}\bibinfo
  {note} {ArXiv: 1104.0210}\BibitemShut {NoStop}%
\bibitem [{\citenamefont {Gemelke}\ \emph {et~al.}(2009)\citenamefont
  {Gemelke}, \citenamefont {Zhang}, \citenamefont {Hung},\ and\ \citenamefont
  {Chin}}]{CL2DMott}%
  \BibitemOpen
  \bibfield  {author} {\bibinfo {author} {\bibfnamefont {N.}~\bibnamefont
  {Gemelke}}, \bibinfo {author} {\bibfnamefont {X.}~\bibnamefont {Zhang}},
  \bibinfo {author} {\bibfnamefont {C.}~\bibnamefont {Hung}}, \ and\ \bibinfo
  {author} {\bibfnamefont {C.}~\bibnamefont {Chin}},\ }\href@noop {} {\bibfield
   {journal} {\bibinfo  {journal} {Nature},\ }\textbf {\bibinfo {volume}
  {460}},\ \bibinfo {pages} {995} (\bibinfo {year} {2009})}\BibitemShut
  {NoStop}%
\bibitem [{dif()}]{diffraction_note}%
  \BibitemOpen
  \href@noop {} {}\bibinfo {note} {We thank Referees for pointing this
  out.}\BibitemShut {Stop}%
\bibitem [{\citenamefont {Gati}\ \emph {et~al.}(2006)\citenamefont {Gati},
  \citenamefont {Albiez}, \citenamefont {Folling}, \citenamefont {Hemmerling},\
  and\ \citenamefont {Oberthaler}}]{boson_Josephson_Junction}%
  \BibitemOpen
  \bibfield  {author} {\bibinfo {author} {\bibfnamefont {R.}~\bibnamefont
  {Gati}}, \bibinfo {author} {\bibfnamefont {M.}~\bibnamefont {Albiez}},
  \bibinfo {author} {\bibfnamefont {J.}~\bibnamefont {Folling}}, \bibinfo
  {author} {\bibfnamefont {B.}~\bibnamefont {Hemmerling}}, \ and\ \bibinfo
  {author} {\bibfnamefont {M.~K.}\ \bibnamefont {Oberthaler}},\ }\href@noop {}
  {\bibfield  {journal} {\bibinfo  {journal} {Appl. Phys. B},\ }\textbf
  {\bibinfo {volume} {82}},\ \bibinfo {pages} {207} (\bibinfo {year}
  {2006})}\BibitemShut {NoStop}%
\bibitem [{\citenamefont {Levy}\ \emph {et~al.}(2007)\citenamefont {Levy},
  \citenamefont {Lahoud},\ and\ \citenamefont {Steinhauser}}]{JJunction}%
  \BibitemOpen
  \bibfield  {author} {\bibinfo {author} {\bibfnamefont {S.}~\bibnamefont
  {Levy}}, \bibinfo {author} {\bibfnamefont {E.}~\bibnamefont {Lahoud}}, \ and\
  \bibinfo {author} {\bibfnamefont {J.}~\bibnamefont {Steinhauser}},\
  }\href@noop {} {\bibfield  {journal} {\bibinfo  {journal} {Nature},\ }\textbf
  {\bibinfo {volume} {449}},\ \bibinfo {pages} {579} (\bibinfo {year}
  {2007})}\BibitemShut {NoStop}%
\bibitem [{Note1()}]{Note1}%
  \BibitemOpen
  \bibinfo {note} {$k_{F}a_{bg}\approx -1.5\times 10^{-3}$ at the trap center
  from Ref.\cite {SU12} so $T_c$ is well below experimentally accessible
  temperatures so far.}\BibitemShut {Stop}%
\bibitem [{\citenamefont {Kinast}\ \emph {et~al.}(2005)\citenamefont {Kinast},
  \citenamefont {Turlapov}, \citenamefont {Thomas}, \citenamefont {Chen},
  \citenamefont {Stajic},\ and\ \citenamefont {Levin}}]{ThermoScience}%
  \BibitemOpen
  \bibfield  {author} {\bibinfo {author} {\bibfnamefont {J.}~\bibnamefont
  {Kinast}}, \bibinfo {author} {\bibfnamefont {A.}~\bibnamefont {Turlapov}},
  \bibinfo {author} {\bibfnamefont {J.~E.}\ \bibnamefont {Thomas}}, \bibinfo
  {author} {\bibfnamefont {Q.~J.}\ \bibnamefont {Chen}}, \bibinfo {author}
  {\bibfnamefont {J.}~\bibnamefont {Stajic}}, \ and\ \bibinfo {author}
  {\bibfnamefont {K.}~\bibnamefont {Levin}},\ }\href@noop {} {\bibfield
  {journal} {\bibinfo  {journal} {Science},\ }\textbf {\bibinfo {volume}
  {307}},\ \bibinfo {pages} {1296} (\bibinfo {year} {2005})}\BibitemShut
  {NoStop}%
\bibitem [{\citenamefont {Zwierlein}\ \emph {et~al.}(2006)\citenamefont
  {Zwierlein}, \citenamefont {Schunck}, \citenamefont {Schirotzek},\ and\
  \citenamefont {Ketterle}}]{Zwierlein2006}%
  \BibitemOpen
  \bibfield  {author} {\bibinfo {author} {\bibfnamefont {M.~W.}\ \bibnamefont
  {Zwierlein}}, \bibinfo {author} {\bibfnamefont {C.~H.}\ \bibnamefont
  {Schunck}}, \bibinfo {author} {\bibfnamefont {A.}~\bibnamefont {Schirotzek}},
  \ and\ \bibinfo {author} {\bibfnamefont {W.}~\bibnamefont {Ketterle}},\
  }\href@noop {} {\bibfield  {journal} {\bibinfo  {journal} {Nature},\ }\textbf
  {\bibinfo {volume} {442}},\ \bibinfo {pages} {54} (\bibinfo {year}
  {2006})}\BibitemShut {NoStop}%
\end{thebibliography}
%

\end{document}